\documentclass[12pt]{article}

\usepackage{verbatim,color,amssymb,amsmath}					
\usepackage{amsthm}		
\usepackage{bm}
\usepackage{dsfont}			
\usepackage{multirow}
\usepackage{setspace}
\usepackage[mathscr]{euscript}
\usepackage{fancyhdr}
\usepackage{enumitem}
\usepackage{graphicx}
\usepackage{geometry}
\usepackage{xcolor}
\usepackage{bbm}

\usepackage{tabularx, booktabs}
\newcolumntype{Y}{>{\centering\arraybackslash}X}
\usepackage{lscape}

\usepackage{subfig}
\usepackage{float}
\usepackage{lineno}

\newtheorem{Thm}{\underline{\bf Theorem}}

\newtheorem*{Proof*}{Proof}

\newtheorem{Prop}{\underline{\bf Proposition}}

\usepackage[ruled,norelsize,vlined]{algorithm2e}
\usepackage[colorlinks,linkcolor=black,citecolor=blue,urlcolor=blue]{hyperref}
\makeatletter
\newcommand*{\rom}[1]{\expandafter\@slowromancap\romannumeral #1@}
\makeatother

\makeindex             

\newcommand{\R}{ {\mathbb{R}} }
\newcommand{\Exp}{ {\mathbb{E}} }

\newcommand{\bmm}{ {\bf m} }

\newcommand{\bx}{ {\bf x} }

\newcommand{\bs}{ {\bf s} }

\newcommand{\by}{ {\bf y} }
\newcommand{\bX}{ {\bf X} }

\newcommand{\bu}{ {\bf u} }

\newcommand{\bH}{ {\bf H} }
\newcommand{\bI}{ {\bf I} }

\newcommand{\bV}{ {\bf V} }

\newcommand{\bW}{ {\bf W} }

\newcommand{\ba}{ {\bf a} }

\newcommand{\bgamma}{ {\boldsymbol \gamma} }

\newcommand{\bLambda}{ {\boldsymbol \Lambda} }

\newcommand{\bGamma}{ {\boldsymbol \Gamma} }
\newcommand{\bDelta}{ {\boldsymbol \Delta} }
\newcommand{\bbeta}{ {\boldsymbol \beta} }
\newcommand{\bmu}{ {\boldsymbol \mu} }

\newcommand{\bSigma}{ {\boldsymbol \Sigma} }

\newcommand{\bOmega}{ {\boldsymbol \Omega} }
\newcommand{\bomega}{ {\boldsymbol \omega} }

\newcommand{\bxi}{ {\boldsymbol \xi} }
\newcommand{\bz}{ {\bf z} }

\newcommand{\post}{ \mbox{\normalfont \tiny pst}}

\newcommand{\bzero}{ {\boldsymbol{0}} }
\newcommand{\bv}{ {\bf v} }

\begin{document}

	\thispagestyle{empty}
	\baselineskip=28pt

		\begin{center}
			{\LARGE{\bf Bayesian Inference for the 
				Multinomial Probit Model under Gaussian Prior Distribution
			}}
		\end{center}
		\vskip 24pt

		\small
		\baselineskip=14pt
		\begin{center}
			Augusto Fasano$^{a,b}$ (augusto.fasano@carloalberto.org)\\
			Giovanni Rebaudo$^{b,c}$ (giovanni.rebaudo@austin.utexas.edu)\\
			Niccol\'o Anceschi$^{b,d}$ (niccolo.anceschi@phd.unibocconi.it)
			
			\vskip 3mm
			$^{a}$Collegio Carlo Alberto,\\
			Piazza Arbarello 8, Torino, Italy\\
			\vskip 4pt 
			$^{b}$Bocconi Institute for Data Science and Analytics, Bocconi University,\\ 
			Via Röntgen 1, 20136 Milan, Italy
			\vskip 4pt 
			$^{c}$Department of Statistics and Data Sciences,
			University of Texas at Austin,\\
			105 East 24th Street D9800, Austin, TX 78712, USA\\	
			\vskip 4pt 
			$^{d}$Department of Decision Sciences, Bocconi University,\\
			via R\"ontgen 1, 20136 Milan, Italy
		\end{center}

		\vskip 24pt 
		
		\begin{center}
			{\Large{\bf Abstract}} 
		\end{center}
			\baselineskip 18pt 
Multinomial probit (\textsc{mnp}) models are fundamental and widely-applied regression models for categorical data.
\cite{fasano2020} proved that the class of unified skew-normal 
distributions is conjugate to several \textsc{mnp} sampling models.
This allows to develop Monte Carlo samplers and accurate variational methods to perform Bayesian inference.
In this paper, we adapt the above-mentioned results for a popular special case: the discrete-choice \textsc{mnp} model 
under zero-mean and independent Gaussian priors.
This allows to obtain simplified expressions for the parameters of the posterior distribution and an alternative derivation for the variational algorithm that gives a novel understanding of the fundamental results in \cite{fasano2020} as well as computational advantages in our special settings.
		
		
		
		
		\pagenumbering{arabic}
\setcounter{page}{0}
\newlength{\gnat}
\setlength{\gnat}{20pt}
\baselineskip=\gnat

\newpage

\section{Introduction}
\label{sec:1}
Multinomial probit (\textsc{mnp}) models constitute a fundamental tool for categorical data regression, thanks to their interpretability and flexibility \cite{greene2003}.
Originally introduced by \cite{hausman1978} to avoid the restrictive assumption of the \textit{independence of irrelevant alternatives} typical of multinomial logit models, such models have faced the growth of many different specifications.
Among them, we consider the Bayesian formulation of the discrete choice \textsc{mnp} model with class-specific effects \cite{stern_1992}, under a zero-mean and independent Gaussian prior for the parameters, adapting the results of \cite{fasano2020} obtained under more general prior specifications and for a wider range of models.
In such a construction, originally developed in the econometrics literature, to each possible choice (or class) $\ell=1,\ldots,L$ that individual $i=1,\ldots,n$ faces, a corresponding random latent utility $z_{i\ell}=\bx_{i}^\intercal \bbeta_\ell + \epsilon_{i\ell}$ is associated, where $\bx_i\in \R^p$ is the covariate vector for observation $i$, $\bbeta_\ell\in\R^p$ is the class-specific vector of the covariate effects and $\boldsymbol{\epsilon}_{i} = (\epsilon_{i1},\ldots,\epsilon_{iL})^\intercal\sim \textsc{n}_L\left(\boldsymbol{0},\bSigma\right)$, independently across units $i=1,\ldots,n$.
Note that the error terms for different alternatives can be correlated, since $\bSigma$, which is assumed to be known, is not necessarily diagonal.
Among all the possible choices $1,\ldots,L$, individual $i$ chooses the one giving her the maximal utility, meaning that $y_i=\ell$ if and only if $z_\ell=\max\{z_1,\ldots,z_L\}$.
Thus, for each $i=1,\ldots,n$, independently
\begin{equation}
\begin{split}
\Pr(y_i=\ell \mid \bbeta_1,\ldots,\bbeta_L,\bx_i)&=\Pr(z_{i\ell}>z_{ik}, \forall k \neq \ell)\\
&=\Pr(\bx^{\intercal}_{i}\bbeta_\ell+ \varepsilon_{i\ell}>\bx^{\intercal}_{i}\bbeta_k+ \varepsilon_{ik}, \forall  k \neq \ell).
\end{split}
\label{eq1}
\end{equation}
Since in \eqref{eq1} only pairwise differences between the parameters matter, we set $\bbeta_L=\boldsymbol{0}$ for identifiability purposes.
In order to perform Bayesian inference, we complete the model by specifying a multivariate Gaussian prior distribution for the vector of parameters $\bbeta=\left(\bbeta_1^\intercal,\ldots,\bbeta_{L-1}^\intercal\right)^\intercal$ centered in zero, with independent and homoscedastic components. Thus,
\begin{equation}
	\bbeta\sim \textsc{n}_{p(L-1)}\left(\boldsymbol{0},\nu^2\bI_{p(L-1)}\right).
	\label{eq2}
\end{equation}
The more general results under a unified skew-normal (\textsc{sun}) prior have been developed in \cite{fasano2020} for a broader class of \textsc{mnp} models.
In fact, it is shown that in that case the posterior belongs again to the class of \textsc{sun} distributions, with updated dimensionalities and parameters. This allows to perform posterior inference via 
i.i.d.\ samples in small-to-moderate $n$ settings, while a blocked variational procedure is developed to avoid the computational bottlenecks that one may encounter in large $n$ scenarios.
The prior specification \eqref{eq2}, however, constitutes a popular choice in case there are no reasons to \textit{a priori} assume any dependence between the parameters or asymmetry in their distribution, and it is worth a separate treatment as it allows the simplification of the expression of some important parameters in the posterior distribution and the derivation of an alternative proof for the variational algorithm that can have computational advantages.
These two important aspects represent the main focus of the present article and will be the focus of Sections \ref{sec:2} and \ref{sec:3} below, while Section \ref{sec:4} is dedicated to the discussion of possible future research directions.

\section{Posterior inference via Monte Carlo samples}
\label{sec:2}
\vspace*{-0.2cm}
Recently, \cite{fasano2020} showed that for a broad class of \textsc{mnp} models, a \textsc{sun} prior distribution leads to a \textsc{sun} posterior distribution, extending the previous conjugacy results derived by \cite{durante2019} for the classical binary probit model.
We specify here results in Section 2.2 in \cite{fasano2020} under the particular Gaussian prior distribution \eqref{eq2}, for the peculiar advantages explained in Section \ref{sec:1}.
Before doing that, we briefly recap the definition and the main properties of the \textsc{sun} distribution.
Further details can be found, for instance, in 
\cite{azzalini_2013}.
A random vector $\bbeta \in \R^q$ has \textsc{sun} distribution, $\bbeta\sim \mbox{\textsc{sun}}_{q,h}(\bxi,\bOmega,\bDelta,\bgamma,\bGamma)$, if its density function $p(\bbeta)$ can be expressed as
\begin{equation*}
p(\bbeta)=\phi_q(\bbeta -\bxi; \bOmega) \frac{\Phi_h[\bgamma+\bDelta^\intercal \bar{\bOmega}^{-1} \bomega^{-1}(\bbeta-\bxi); \bGamma{-}\bDelta^{\intercal}\bar{\bOmega}^{-1}\bDelta ]}{\Phi_h(\bgamma;\bGamma)},
\end{equation*}
where the covariance matrix $\bOmega$ of the Gaussian density $\phi_q(\bbeta -\bxi; \bOmega)$ can be decomposed as $\bOmega=\bomega \bar{\bOmega} \bomega$, i.e.\ by rescaling the correlation matrix $\bar{\bOmega}$ via the diagonal scale matrix $ \bomega=(\bOmega \odot {\bf I}_q)^{1/2}$, with $\odot$ denoting the element-wise Hadamard product.
Moreover, $\Phi_h(\bu;\bW)$ denotes the cumulative distribution function of a $\textsc{n}_h(\boldsymbol{0},\bW)$ evaluated at $\bu$.
The following additive characterization constitutes a fundamental property to further understand the role of the parameters and to develop an i.i.d.\ sampler. If $\bbeta \sim \textsc{sun}_{q,h}(\bxi, \bOmega, \bDelta, \bgamma, \bGamma)$, then $\bbeta\stackrel{\mbox{\small d}}{=}\bxi+\bomega(\bV_0+\bDelta\bGamma^{-1}\bV_1)$, with
\begin{equation*}
	 \bV_0 \sim\textsc{n}_q(\boldsymbol{0},\bar{\bOmega}-\bDelta\bGamma^{-1}\bDelta^\intercal), \ \bV_1\sim\textsc{tn}_h(\boldsymbol{0},\bGamma; A_{-\bgamma}), 
	\label{eq3}
\end{equation*}
 where $A_{-\bgamma}=\left\{\ba\in \R^h\colon a_i \ge -\gamma_i\ \forall i \right\}$ and $\textsc{tn}_{h}(\bmm,\bW;A)$ denotes the $h$-variate normal distribution with mean $\bmm$ and covariance matrix $\bW$, truncated in the region $A$.

In order to derive the \textsc{sun} posterior distribution of $\bbeta$ for model \eqref{eq1}-\eqref{eq2}, we explicitly write the likelihood expression for the observed responses $\by=(y_1, \ldots, y_n)^\intercal$. 

\begin{Prop}[Proposition 2 in \cite{fasano2020}]
	For each $\ell=1, \ldots, L$, denote with $\bv_\ell$ the $L \times 1$ vector with value $1$ in position $l$ and $0$ elsewhere, and with $\bV_{[-\ell]}$ the $(L-1) \times L$ matrix whose rows are obtained by stacking vectors $(\bv_k-\bv_\ell)^{\intercal}$, for $k\ne \ell$.
	Finally, define $\bX_{i[-\ell]}=-\bar{\bV}_{[-\ell]} \otimes \bx_i^\intercal$, where  $\bar{\bV}_{[-\ell]}$ is the $(L-1) \times (L-1)$ matrix obtained by removing the $L$-th column from $\bV_{[-\ell]}$ and  $\otimes$ denotes the Kronecker product.
	\begin{equation*}
		%
		p(\by \mid \bbeta, \bX)=\prod_{i=1}^n \Phi_{L-1}(\bX_{i[-y_i]} \bbeta; \bV_{[-y_i]} \bSigma \bV^{\intercal}_{[-y_i]})= \Phi_{n(L-1)}(\bar{\bX} \bbeta; \bLambda),
		\label{eq4}
	\end{equation*}
	where $\bar{\bX}$ is an $n(L-1) \times p(L-1)$ block matrix with $(L-1) \times p(L-1)$ row blocks $\bar{\bX}_{[i]}=\bX_{i[-y_i]}$, whereas $\bLambda$ denotes an $n(L-1) \times n(L-1)$ block-diagonal covariance matrix with $(L-1) \times (L-1)$ diagonal blocks $\bLambda_{[ii]}=\bV_{[-y_i]} \bSigma \bV^{\intercal}_{[-y_i]}$, for $i=1, \ldots, n$.
	\label{prop2}
\end{Prop}
By combining the prior specification \eqref{eq2} with the likelihood \eqref{eq4}, the following closed-form expression for the posterior distribution of $\bbeta$ is obtained as a direct consequence of Theorem 1 in \cite{fasano2020} adapted to the special case described in Section \ref{sec:1}.

\begin{Thm}[from Theorem 1 in \cite{fasano2020}]
	Under model \eqref{eq1}-\eqref{eq2}, the posterior density is
	\begin{equation}
		(\bbeta \mid \by, \bX)\sim \mbox{\textsc{sun}}_{p(L-1),n(L-1)}(\bzero,\bOmega_{\post},\bDelta_{\post},\bzero ,\bGamma_{\post})
		\label{eq5}
	\end{equation}
	with $\bOmega_{\post}=\nu^2\bI_{p(L-1)}$, $\bDelta_{\post}=\nu \bar{\bX}^{\intercal}\bs^{-1}$ and $\bGamma_{\post}$ is an $n(L-1) \times n(L-1)$ correlation matrix $\bGamma_{\post}=\bs^{-1}(\nu^2\bar{\bX}\bar{\bX}^{\intercal}+\bLambda)\bs^{-1}$, where $\bs=[(\nu^2 \bar{\bX}\bar{\bX}^{\intercal}+\bLambda)\odot \bI_{n(L-1)}]^{1/2}$.
	\label{teo1}
\end{Thm}
Theorem \ref{teo1} and the additive representation of the \textsc{sun} \eqref{eq3} allow to develop an i.i.d.\ sampler of the posterior of $\bbeta$ as described in Algorithm \ref{algo1}.

\begin{algorithm*}[t]
	\small
	\caption{Strategy to sample from the \textsc{sun} posterior in Theorem \ref{teo1}} 
	\For{i=1,\ldots,N}{
		\mbox{{\bf [1]} Sample $\bV^{(i)}_0 \sim\textsc{n}_{p(L-1)}(\boldsymbol{0},\bI_{p(L-1)}-\bDelta_{\post}\bGamma_{\post}^{-1}\bDelta_{\post}^\intercal)$ [in \texttt{R} use the function \texttt{rmvnorm}]} \\
		\mbox{{\bf [2]} Sample  $\bV^{(i)}_1 \sim\textsc{tn}_{n(L-1)}(\boldsymbol{0},\bGamma_{\post}; [0,\infty)^{n(L-1)})$ [in \texttt{R} use the function \texttt{rtmvnorm}]}  \\
		{\bf [3]} Compute $\bbeta^{(i)}=\nu (\bV^{(i)}_0+\bDelta_{\post}\bGamma_{\post}^{-1}\bV^{(i)}_1)$ \\
	}
	\mbox{{\bf Output:} i.i.d.\ samples $\bbeta^{(1)}, \ldots, \bbeta^{(N)}$ from the posterior distribution in Theorem \ref{teo1}.} 
	\label{algo1}
\end{algorithm*}

\section{Partially-factorized blocked mean-field approximation (PFM-B)}
\label{sec:3}

Basic manipulations of the posterior \eqref{eq5} show that $p(\bbeta\mid\by,\bX)\propto p(\bbeta) \cdot \Pr[\bar{\bz} > \boldsymbol{0}\mid \bbeta, \bar{\bX}]$, where $\bar{\bz}=\left(\bar{\bz}_1,\ldots,\bar{\bz}_n\right)^\intercal\in \R^{n(L-1)}$ and $\bar{\bz}\mid\bbeta,\bar{\bX} \sim \textsc{n}_{n(L-1)}\left(\bar{\bX}\bbeta,\bLambda\right)$.
Thus, $p(\bbeta\mid\by,\bX)$ can be seen as the marginal posterior distribution of the dual model 
\begin{subequations}
	\begin{align}
		\bbeta &\sim \textsc{n}_{p(L-1)}\left(\boldsymbol{0},\nu^2\bI_{p(L-1)}\right) \label{eq6a}\\		
		\bar{\bz}_i\mid\bbeta,\bar{\bX}&\overset{ind}{\sim}\textsc{n}_{L-1}\left(\bar{\bX}_{[i]}\bbeta,\bLambda_{[ii]}\right),\quad i=1,\ldots,n, \label{eq6b}\\		
		\bar{\by}_i &= \mathbbm{1}\left[\bar{\bz}_i > \boldsymbol{0}\right],\quad\quad\quad\quad\quad i=1,\ldots,n, 
		 \label{eq6c}	 		
	\end{align} \label{eq6}
\end{subequations}
in which one observes $\bar{\by}_i=\left(1,\ldots,1\right)^\intercal\in \R^{L-1}$ for $i=1,\ldots,n$ and  $\mathbbm{1}[\cdot]$ in \eqref{eq6c} is intended component-wise.
Thus, it holds $p(\bbeta\mid \by,\bX)=\int_{\R^{n(L-1)}} p(\bbeta,\bar{\bz}\mid \bar{\by},\bar{\bX}) d\bar{\bz}$, with $\bar{\by}=\left(\bar{\by}_1,\ldots,\bar{\by}_n\right)^\intercal$ an $n(L-1)$ vector of ones.
Since direct sampling from $p(\bbeta\mid \by, \bX)$ can run into computational issues when $n(L-1)$ is large due to step 2 in Algorithm~\ref{algo1}, one can resort to variational methods, see, e.g., \cite{blei_2017}, to compute the `best' possible approximating joint density $q^*(\bbeta,\bar{\bz})$ in a given class of tractable density functions $\mathcal{Q}$.
This optimal solution is the minimizer of the Kullback-Leibler divergence \cite{Kullback_1951} $\textsc{kl}\left[q(\bbeta,\bar{\bz}) \mid \mid  p(\bbeta,\bar{\bz}\mid \bar{\by},\bar{\bX})\right]=\Exp_{q(\bbeta,\bar{\bz})}\left[\log\left(q(\bbeta,\bar{\bz})/p(\bbeta,\bar{\bz}\mid \bar{\by},\bar{\bX})\right)\right]$. 

In order to get a tractable approximation, but retain as much structure of the posterior distribution as possible, following \cite{fasano2020}, we take $\mathcal{Q} = \mathcal{Q}_{\textsc{pfm-b}}=\{ q(\bbeta, \bar{\bz}): q(\bbeta, \bar{\bz})=q(\bbeta \mid\bar{\bz} ) \prod_{i=1}^{N} q(\bar{\bz}_i)\}$.
This class of densities leverages on the partially-factorized mean field approximation developed for the classical probit model in \cite{fasano2019}, but allows to maintain the intra-correlations between the components of the $\bar{\bz}_i$'s, $i=1,\ldots,n$, while enforcing the inter-correlations between them to be zero.
A first result about the optimal approximating density is obtained by the chain rule for the \textsc{kl} divergence: $\textsc{kl}[q(\bbeta, \bar{\bz})||p(\bbeta, \bar{\bz} \mid \bar{\by}, \bar{\bX}) ]=\textsc{kl}[q(\bar{\bz})||p(\bar{\bz} \mid \bar{\by}, \bar{\bX}) ]+\mathbb{E}_{q(\bar{\bz})}\{\textsc{kl}[q(\bbeta \mid\bar{\bz} )||p(\bbeta \mid \bar{\bz},\bar{\by}, \bar{\bX})] \}$.
Thus, whatever is the value for $q(\bar{\bz})$, the second summand is zero, and hence minimal, if and only if, calling $\bV=(\nu^{-2}\bI_{p(L-1)} + \bar{\bX}^\intercal\bLambda^{-1}\bar{\bX})^{-1}$,

\begin{equation*} 
		q(\bbeta \mid\bar{\bz} ) = q^*(\bbeta \mid\bar{\bz} ) = p(\bbeta \mid \bar{\bz},\bar{\by}, \bar{\bX})= \phi_{p(L-1)}(\bbeta - \bV\bar{\bX}^\intercal \bLambda^{-1} \bar{\bz}; \bV),
\label{eq7}
\end{equation*}
which follows by standard properties of multivariate normals applied to model \eqref{eq6}.
This means that in order to find the \textsc{kl} minimizer $q^*(\bbeta, \bar{\bz})=q^*(\bbeta \mid\bar{\bz} )\prod_{i=1}^{N} q^*(\bar{\bz}_i)$  in $\mathcal{Q}_{\textsc{pfm-b}}$, we just have to find $q^*(\bar{\bz})=\prod_{i=1}^{N} q^*(\bar{\bz}_i)$ minimizing $\textsc{kl}[q(\bar{\bz})||p(\bar{\bz} \mid \bar{\by}, \bar{\bX})]$.
See \cite{fasano2020} for further details.
Due to the imposed factorization for $q^*\left(\bar{\bz}\right)$, which takes the name of `mean-field approximation', the desired solution is known to satisfy the following mean field equations (see \cite{blei_2017} for additional details):

\begin{equation*}
	\log q^*(\bar{\bz}_i) \propto  \mathbb{E}_{q^*(\bar{\bz}_{-i})}[\log p(\bar{\bz}_i\mid \bar{\bz}_{-i}, \bar{\by}, \bar{\bX})],\quad i=1,\ldots,n,
\end{equation*}
where the expectation is taken with respect to the distribution of all the $\bar{\bz}_j$ other than $\bar{\bz}_i$. 
Here, we show how this solution can be obtained by exploiting the dual hierarchical model \eqref{eq6}. This gives further intuition and a broader understanding of the procedure.
Moreover, besides specifying to the current setting the results derived in \cite{fasano2020} under more general constructions, it allows computational gains due to the simplification of certain parameters, above all the covariance matrices in $q^*(\bar{\bz}_i)$.

\setlength{\textfloatsep}{10pt}
\begin{algorithm*}[t]
	\small
	\caption{\mbox{\textsc{cavi} algorithm for ${q}^*(\bbeta,\bar{\bz})={q}^*(\bbeta\mid\bar{\bz}) \prod_{i=1}^n {q}^*(\bar{\bz}_i)$}}
	\label{algo2}

	\mbox{{\bf [1]} For each $i=1, \ldots, n$, set $\bSigma_i^* = \left(\bLambda_{[i,i]}^{-1} -\bH_{[i,i]} \right)^{-1}$ and initialize  $\Exp_{q^{0}\left(\bar{\bz}_i\right)}[\bar{\bz}_i]\in\R^{(L-1)}$}\\

	\begingroup
	{{\bf [2]}      \For(){$t$ \mbox{from} $1$ until convergence}
		{

			\For(){$i$ \mbox{from} $1$ \mbox{to} $n$}
			{ 
				\mbox{{\bf [2.1]} Set $\bmu^{(t)}_i=\bSigma_i^*\bH_{[i,-i]}(\Exp_{q^{(t)}}[\bar{\bz}_1]^\intercal, \ldots, \Exp_{q^{(t)}}[\bar{\bz}_{i-1}]^\intercal, \Exp_{q^{(t-1)}}[\bar{\bz}_{i+1}]^\intercal, \ldots,  \Exp_{q^{(t-1)}}[\bar{\bz}_{n}]^\intercal)^{\intercal}$}\\
				\mbox{{\bf [2.2]} Compute $\Exp_{q^{(t)}}[\bar{\bz}_i]$ with $\bar{\bz}_i\sim \textsc{tn}_{L-1}\left(\bmu^{(t)}_i, \bSigma_i^*; [0,\infty)^{L-1} \right)$ [in \texttt{R} use  \texttt{MomTrunc()}]}
			}
		}
		
	}
	{\bf [3]} Set $q^*(\bar{\bz}_i)=q^{(t)}(\bar{\bz}_i)$ for $i=1,\ldots,n$
		\vspace{5pt}\\
	{{\bf [4]} Set $q^*(\bbeta \mid\bar{\bz} ) = \phi_{p(L-1)}(\bbeta - \bV\bar{\bX}^\intercal \bLambda^{-1} \bar{\bz}; \bV)$
		\vspace{5pt}
		
		{\bf Output:} ${q}^*(\bbeta,\bar{\bz})=q^*(\bbeta \mid\bar{\bz} ) \prod_{i=1}^{N} q^*(\bar{\bz}_i)$}
	\endgroup
\end{algorithm*}

First, by marginalizing out $\bbeta$ in \eqref{eq6b}, we get $\bar{\bz}\mid \bar{\bX}\sim \textsc{n}_{n(L-1)}\left(\boldsymbol{0},\bLambda + \nu^2 \bar{\bX}\bar{\bX}^\intercal\right)$, thus $\bar{\bz}\mid \bar{\by},\bar{\bX}\sim \textsc{tn}_{n(L-1)}\big(\boldsymbol{0},\bLambda + \nu^2 \bar{\bX}\bar{\bX}^\intercal;[0,\infty)^{n(L-1)}\big)$.
Taking $\bV$ as above and $\bH = \bLambda^{-1} \bar{\bX}\bV\bar{\bX}^\intercal\bLambda^{-1}$, by Woodbury's identity it holds $(\bLambda + \nu^2\bar{\bX}\bar{\bX}^\intercal)^{-1} = \bLambda^{-1} - \bH$ \mbox{and so}
\begin{equation*}
	\begin{split}
		p(\bar{\bz}_i\mid \bar{\bz}_{-i}, \bar{\by}) &\propto \exp\big\{-0.5 \, \bar{\bz}^\intercal (\bLambda + \nu^2\bar{\bX}\bar{\bX}^\intercal)^{-1} \bar{\bz} \big\}\mathbbm{1}[\bar{\bz}_i > \boldsymbol{0}]\\
		&\propto \exp\big\{-0.5 \, \bar{\bz}^\intercal ( \bLambda^{-1} - \bH) \bar{\bz} \big\}\mathbbm{1}[\bar{\bz}_i > \boldsymbol{0}]\\
		&\propto \exp\big\{-0.5 \, \bar{\bz}^\intercal ( \bLambda_{[i,i]}^{-1} -\bH_{[i,i]} ) \bar{\bz} +  \bar{\bz}_i^\intercal \bH_{[i,-i]} \bar{\bz}_{-i} \big\}\mathbbm{1}[\bar{\bz}_i > \boldsymbol{0}],\\
	\end{split}
\end{equation*}
from which we get
\begin{equation*}
		q^*(\bar{\bz}_{i}) 
		\propto \exp\left\{-\frac{1}{2} \bar{\bz}^\intercal\big(\bLambda_{[i,i]}^{-1} -\bH_{[i,i]} \big) \bar{\bz} + \bar{\bz}_i^\intercal \bH_{[i,-i]} \mathbb{E}_{q^*(\bar{\bz}_{-i})}[ \bar{\bz}_{-i}]\right\}\mathbbm{1}[\bar{\bz}_i > \boldsymbol{0}],
\end{equation*}
which shows that $q^*(\bar{\bz}_{i})$ is the density of a  normal random variable with parameters $\bmu_i^* = \bSigma_i^*\bH_{[i,-i]} \mathbb{E}_{q^*(\bar{\bz}_{-i})}[ \bar{\bz}_{-i}]$ and $\bSigma_i^* = \left(\bLambda_{[i,i]}^{-1} -\bH_{[i,i]} \right)^{-1}$, truncated above zero.
In order to obtain in practice the optimal \textsc{pfm-b} variational solution, one can resort to standard \textsc{cavi} algorithms (see, e.g., \cite{blei_2017}), as shown in detail in Algorithm \ref{algo2}.
It is worth noting that, differently from Algorithm \ref{algo1}, in Algorithm \ref{algo2} we only have to deal with expectations of $(L-1)$-variate truncated normals, significantly reducing the computational burden.
After ${q}^*(\bbeta,\bar{\bz})$ has been computed, approximate posterior moments for $\bbeta$ can be easily obtained leveraging on the law of total expectation 
 as
\begin{equation*}
		\Exp_{q^*(\bbeta)}(\bbeta)=\bV\bar{\bX}^\intercal \bLambda^{-1} \Exp_{q^*(\bar{\bz})}[\bar{\bz}], \quad
		\mbox{var}_{q^*(\bbeta)}(\bbeta)=\bV+\bV\bar{\bX}^\intercal \bLambda^{-1}\mbox{var}_{q^*(\bar{\bz})}(\bar{\bz})\bLambda^{-1}\bar{\bX}\bV,
\end{equation*}
while more complicated functionals can be computed with i.i.d.\ sampling, which, due to the particular block-diagonal structure of the resulting covariance matrix of the multivariate truncated normal distribution $q^*(\bar{\bz})$, would require sampling only from multivariate truncated normals of dimension $L-1$.

\section{Discussion}
\label{sec:4}
This article provides a novel derivation for the \textsc{pfm-b} variational method for a relevant class of \textsc{mnp} models with independent Gaussian priors. 
As shown in Section \ref{sec:3}, this provides a novel understanding of the results in \cite{fasano2020} as well as computational advantages in our special settings.
Future works include deriving results for the posterior 
also for the dynamic \textsc{mnp}, which allows to model sequential decisions in the time series framework.
In such a way, we plan to extend closed-form expressions, and the related samplers, for the filtering, predictive and smoothing distributions of multivariate dynamic probit models for binary time series in \cite{fasano2021}.
Finally, the variational Bayes approach in Section \ref{sec:3} can be generalized to perform inference for the smoothing distribution of the dynamic \textsc{mnp} extending the results in \cite{fasano2021b}.

\end{document}